\documentclass[useAMS,usenatbib,letters]{mn2e}

\usepackage{epsfig}

\title[Counting individual galaxies from deep 24 $\mu$m Spitzer surveys]
{Counting individual galaxies from deep 24 $\mu$m Spitzer surveys: beyond
the confusion limit}
\author[G. Rodighiero et al.]
{G. Rodighiero$^{1}$, C. Lari$^{2}$,
  F. Pozzi$^{3,}$$^{4}$, C. Gruppioni$^{4}$, D. Fadda$^{5}$,  A. Franceschini$^{1}$,
 \newauthor C. Lonsdale$^{6}$, J. Surace$^{6}$, D. Shupe$^{6}$ and F. Fang$^{6}$\\
$^{1}$Dipartimento di Astronomia, Universita' di Padova, Vicolo dell'Osservatorio 5, 351
22 Padova, Italy\\
$^{2}$Istituto di Radioastronomia del CNR, via Gobetti 101, I-40129 Bologna, Italy\\
$^{3}$Dipartimento di Astronomia, Universita' di Bologna, via Ranzani 1, I-40127 Bologna
, Italy\\
$^{4}$Istituto nazionale di Astrofisica, Osservatorio Astronomico di Bologna, via Ranzani
, I-40127 Bologna, Italy\\
$^{5}$Spitzer Science Center, California Institute of
Technology, MC 220-6, Pasadena, CA 91125\\
$^{6}$Infrared Processing and Analysis Center, California Institute of
Technology, MC 100-22, Pasadena, CA 91125}

\begin{document}

\maketitle

\begin{abstract}
We address the question of how to deal with confusion limited surveys in the
mid-infrared domain by using informations from higher frequency observations
over the same sky regions. Such informations, once applied to apparently extended
mid-infrared sources, which are indeed ``blends'' of two or more different
sources, allow us to disentangle the single counterparts and to split the
measured flux density into different components. We present the application 
of this method to the 24 $\mu$m {\em Spitzer} archival data in the GOODS-EN1
test field, where apparently extended, ``blended'' sources constitute about
20\% of a reliable sample of 983 sources detected above the 5$\sigma$ threshold
down to 23 $\mu$Jy. As higher frequency data-set we have considered the 
public IRAC images and catalogues on the same field. 
We show that the 24 $\mu$m sample is almost unbiased down to $\sim$40 $\mu$Jy
and the careful application of the deblending procedure does not
require any statistical completeness correction (at least at the flux level considered).
This is probed by direct comparison of our results with those of Chary
et al. (2004), who analysed the same data-set through extensive Monte
Carlo simulations. The deblending procedure reduces of about 30\% the confusion limit 
of the MIPS 24 $\mu$m survey, allowing one to obtain reliable source counts down to
$\sim$ 40$\mu$Jy. The extrapolation of the source counts down to fainter fluxes 
suggests that our 24 $\mu$m sample is able to resolve
$\sim$62\% of the cosmic background down to a flux level of 38 $\mu$Jy.
\end{abstract}

\maketitle

\section{introduction}
Cosmological surveys in the Mid-Infrared (MIR) / Far-Infrared (FIR) spectral range
reveal a substantial population of strongly evolving dust-enshrouded galaxies at
intermediate redshifts (e.g. Elbaz et al. 1999; Franceschini et al. 2003). These
galaxies are responsible for most of the cosmic infrared background detected 
at these wavelengths (CIRB, Hauser \& Dwek 2001). Although a significant
fraction of the CIRB is already resolved into discrete sources by deep infrared surveys
(i.e. ISOCAM have resolved about 60-70\% of the CIRB at 15 $\mu$m), to resolve
the whole observed background light we should reach fainter flux limits.

However, the main limitation of deep surveys performed in the MIR/FIR domain is
confusion due to extragalactic sources. In fact, the large number of distant galaxies
in deep extragalactic surveys produces a high density of sources with respect to
the instrument beam size. This makes more than one source responsible for the
measured flux density at the fainter limits, therefore producing wrong source counts
(i.e. higher than real at intermediate flux densities due to source ``blending'')
and limiting the effective survey sensitivity.

As recently suggested by Dole et al. (2005), we will show that is
possible to deal with this problem, significantly lowering
the confusion limit in deep extragalactic surveys by making use of higher frequency
observations on the same region of the sky. In particular, data obtained with the
Infrared Array Camera (IRAC; Fazio et al. 2004) on board of {\em Spitzer} in the
3.6 to 8.0 $\mu$m wave-bands are extremely helpful for measuring fluxes of
faint sources and reducing the confusion noise in surveys performed with the {\em Spitzer}
Multiband Imaging Photometer (MIPS; Rieke et al. 2004) at 24, 70 or 160 $\mu$m. 
Through IRAC data, we have developed an efficient ``deblending''
technique that allows us
to accurately measure the flux density of sources detected by MIPS below the nominal 
confusion limit and therefore to significantly lower this limit (of about 30--50\%).
The application of this technique to the GOODS ELAIS-N1 (EN1) test field is
presented in this paper.

The paper is structured as follows: in section 2 we present the observations
in the GOODS-EN1 field; in section 3 we describe the 24 $\mu$m data
processing and we present the 24 $\mu$m catalogue, while in section 4
we discuss the ``deblending'' technique; in section 5 we show the
number counts; in section 6 we present our conclusions.

Throughout this paper we will assume $H_0 = 75$ km s$^{-1}$ Mpc$^{-1}$,
$\Omega_m = 0.3$ and $\Omega_{\Lambda} = 0.7$.

\section{Observations}
\label{obs}
The EN1 field is one of the best areas, among the currently available ones,
for investigating the effects of source confusion in MIR surveys.
Deep MIPS 24-$\mu$m observations in this field cover an
area of  $\sim$ 185 arcmin$^2$ (centred at 16:09:20 +54:57:00)
as part of the Great Observatories Origins Deep Survey
(GOODS) Science Verification program of {\em Spitzer}.
The field was observed in photometry mode. Two different AORs were
executed in cluster mode using the offset position in order to
symmetrically observe one side of the field. 
The integration time per detector pixel varies from $\sim$2000 s, in
the lowest signal to noise area, up to $\sim$4600 s in the deepest regions. 
Details on the MIPS 24-$\mu$m data processing, source extraction and 
photometry will be given in the next section.

Although during the same verification observing campaign complementary
deep NIR observations of EN1 have been obtained with IRAC with the 3.6, 4.5, 5.8 and 8.0
$\mu$m channels, the combined IRAC GOODS maps do not perfectly overlap
the whole MIPS GOODS image, thus leaving a fraction of the MIR map uncovered by NIR data.
However, a shallower and more extended observation of EN1 with the
four IRAC channels has been performed as part of 
the SWIRE Legacy programme (Lonsdale et al. 2003), covering a much wider area 
(4$\times$4 deg$^2$), and thus homogeneously overlapping the deeper GOODS verification 
MIPS data. We therefore used the SWIRE map at 3.6 $\mu$m to have a complete NIR coverage of our MIPS image. 

Optical imaging of the GOODS EN1 field is provided by the First Look
Survey (FLS) observations (Fadda et al., 2004). 
The R-band images were obtained using the Mosaic-1 camera on the 4 m
Mayall Telescope of the Kitt Peak National Observatory.
The images  reach a median 5$\sigma$
depth limiting magnitude of R = 25.5 (Vega) as measured within a 1.35 
FWHM aperture, for which the signal-to-noise ratio is maximal.
\begin{figure*}
\centerline{
\psfig{file=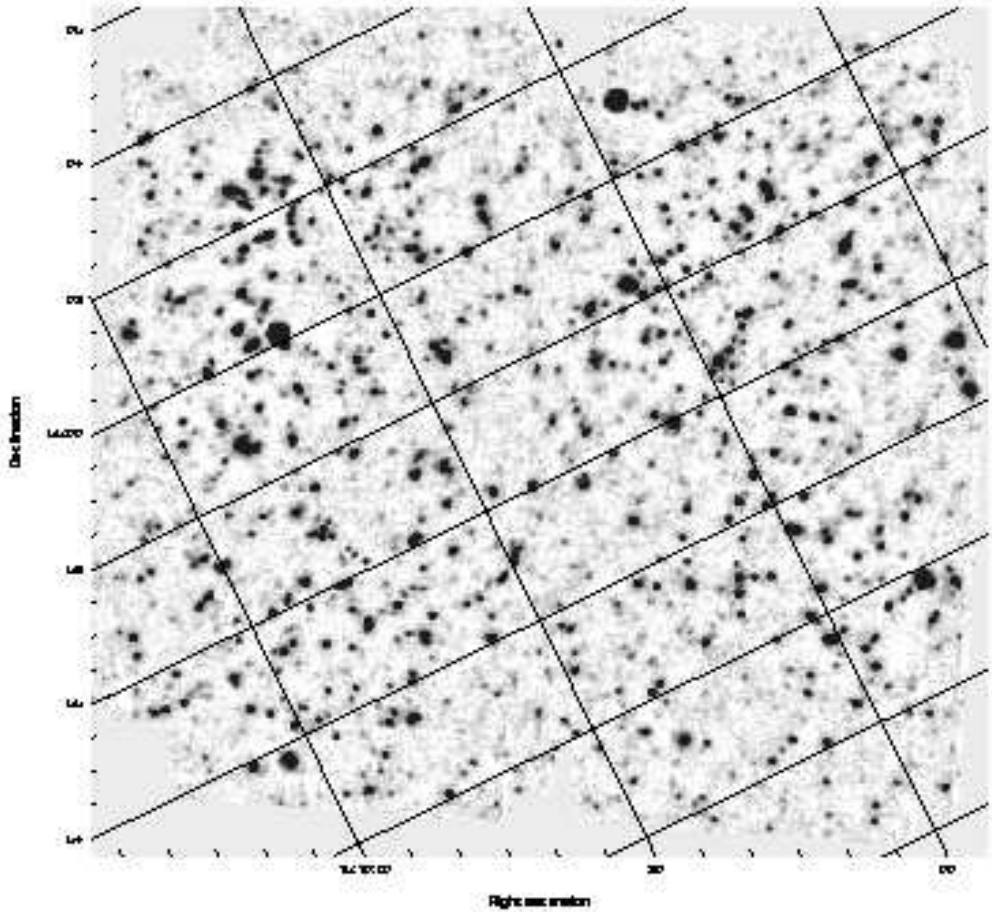,width=17cm}
}
\caption{Signal to noise map of the deep EN1 field observed by MIPS.}
\label{map1}
\end{figure*}

\subsection{IRAC data processing}
The reduction of GOODS data in the IRAC bands has been performed by starting from
the basic calibrated data (BCD) obtained from the {\em Spitzer} archive. We
have applied an additive correction factor to each BCD frame in order to remove
the median background. We have then processed and mosaiced together all the corrected
BCDs with the {\em Spitzer} Mopex package\footnote{See
  http://ssc.spitzer.caltech.edu/postbcd.)} version distributed on
June 2004 by the  Spitzer Science Center (SSC).  
We refer to Lonsdale et al. 
(2004) for a complete description of the observation strategy and data analysis of SWIRE.  

The IRAC source extraction has been performed with SExtractor (Bertin \& Arnouts, 1996), both
on SWIRE and GOODS maps. For point-like sources, we computed the
fluxes within a 6 
arcsec diameter aperture. We have then applied a correction factor derived from the stars in the
images to obtain the total fluxes (1.14).
In the case of extended sources, we used Kron like magnitudes 
(AUTO$\_$MAG output parameter in SExtractor).
The IRAC photometry has been basically used to remove stars in the 24 $\mu$m catalogue.

\section{MIPS 24 $\mu$m data processing}
\label{reduc}
\subsection{Mosaic generation and manipulation}
We started the data analysis of the 24 $\mu$m data in the GOODS EN1 field 
from the archival BCD products. These have been pre-processed using version S10.0.3 of
the SSC pipeline (see Spitzer Observer Manual).
Firstly, we have corrected each single BCD frame by computing a residual
median flat-field which depends on the scan mirror position. 
Such flat-field has been built from the data.

By checking the temporal histories of each pixel we observed a
non-linear response of the detector, that does not instantly reach the 
stabilization level. This produces a transient on the observed  
time sequence of MIPS pixels. We corrected this effect by applying a
linear fit to the temporal sequence of each pixel.
As stabilization level we choose the value of the final exposure frame. 
With this procedure we produced background subtracted and flat-fielded
frames that were co-added using the SSC software Mopex to
obtain a mosaic with half the original pixel scale (1.2 arcsec). 

The projection algorithm applies a linear interpolation that accounts for the distortion 
corrections. Bad pixels are masked within the pipeline and cosmic rays are flagged
using a multiframe temporal outlier detection. 
The signal-to-noise map of the EN1 GOODS field observed by MIPS is shown in Figure \ref{map1}.
\begin{figure}
\psfig{file=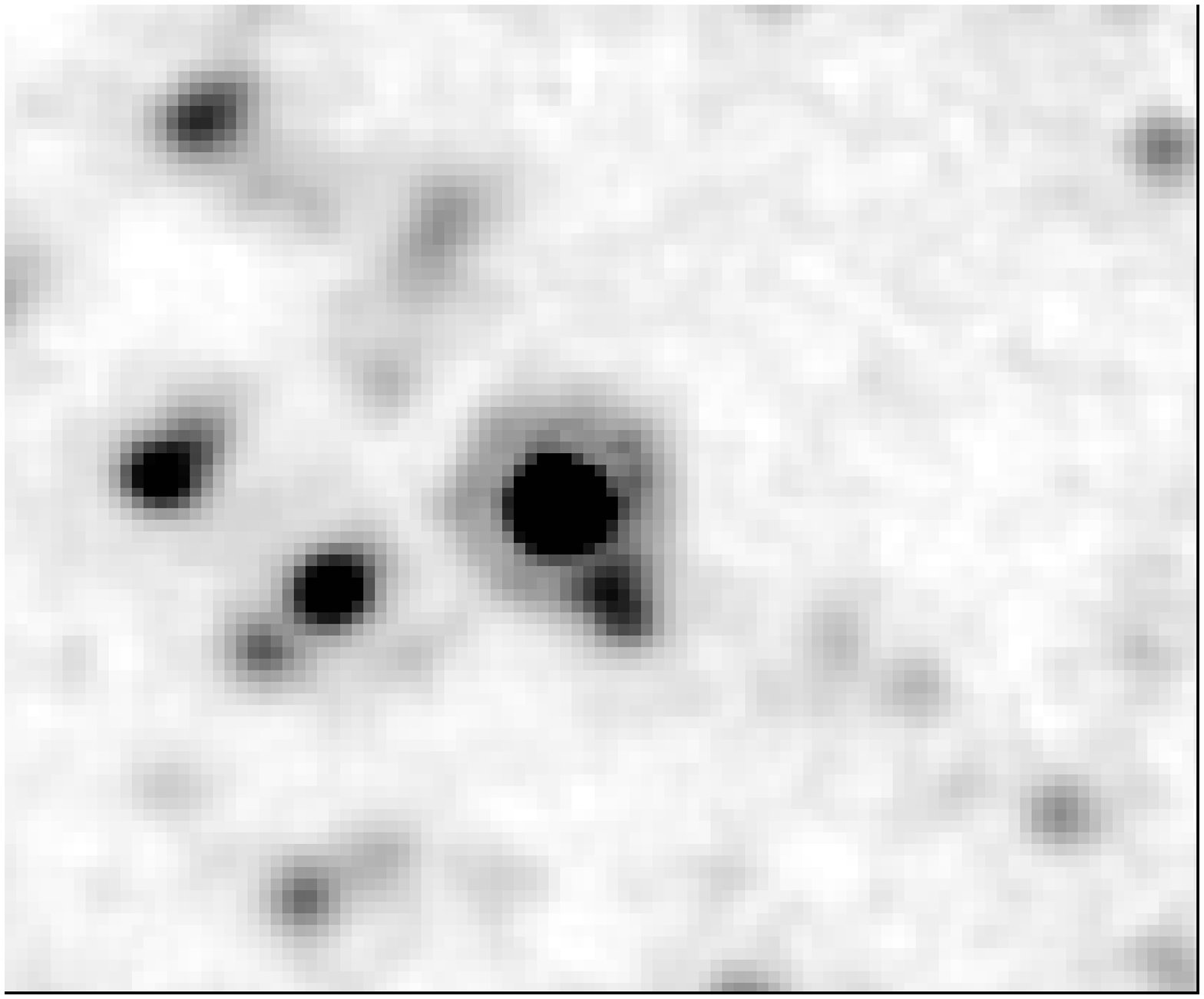,width=8cm}
\psfig{file=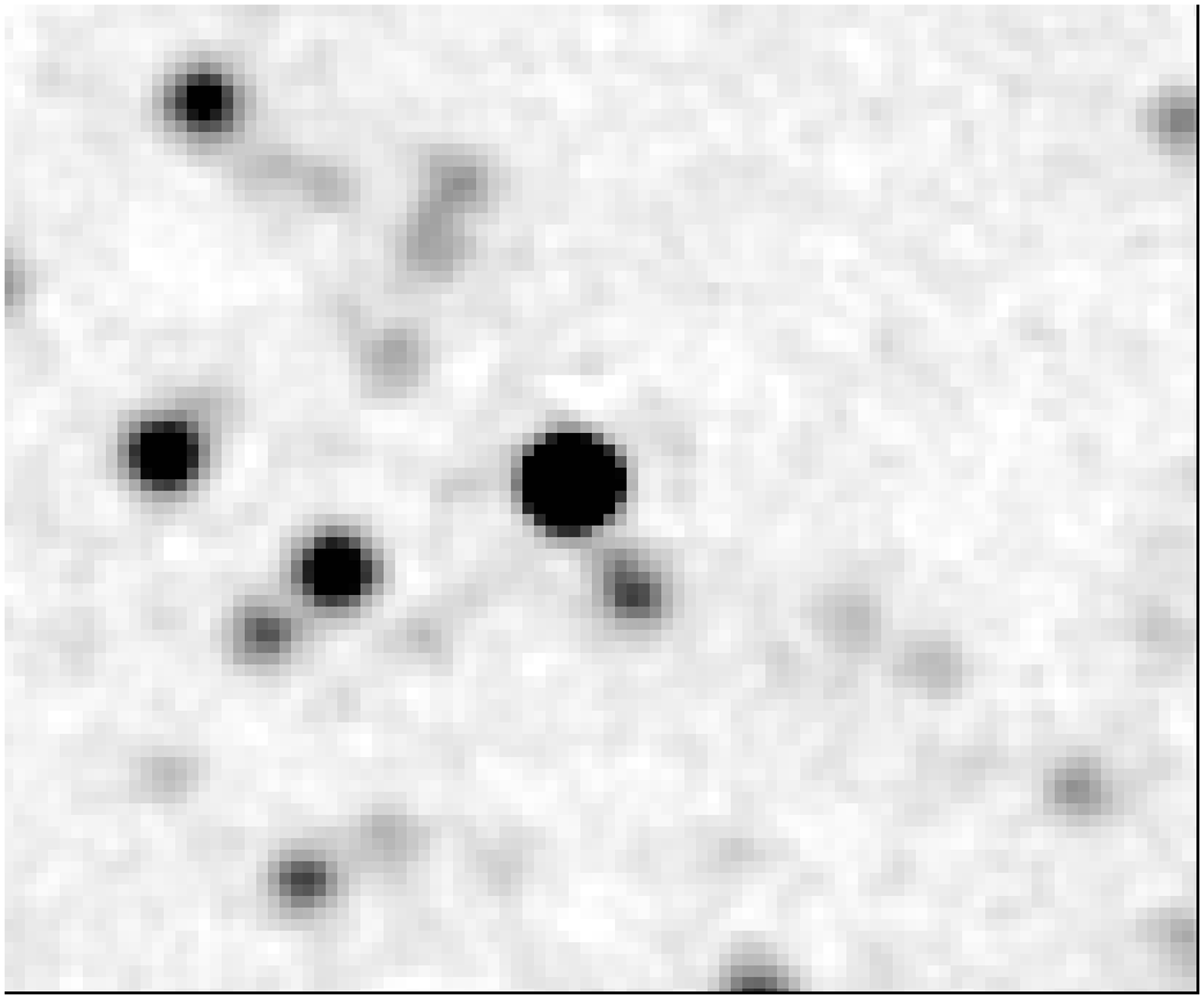,width=8cm}
\caption{Example of the effects of applying the CLEAN task to the
flux map. 
We report a zoom of 2.2$\times$1.4 square arcminutes into the map presented
in Figure \ref{map1}. The upper panel shows the original flux map, in
the lower panel the corresponding cleaned map shows that the adopted
approach is efficient in deconvolving many close sources, providing
a reconstructed map where the confusion due to first Airy ring.
}
\label{map2}
\end{figure}

In order to remove the Airy rings around sources
we have deconvolved the 24 $\mu$m image using the CLEAN task
(typically used for analysing radio data), which is part of the NRAO AIPS
reduction package.
The CLEAN algorithm iteratively finds and substracts positive features on the original map 
until the standard deviation of the residual image is lower than the
noise level. 
The CLEAN map is then constructed by adding to the residual image the
CLEAN components found, convolved with a Gaussian of FWHM 
that is equal to that of the MIPS 24 $\mu$m PSF (we computed a value
of 5.5 arcsec by fitting the PSF provided with the used version of Mopex). This gaussian beam is
normalized to one in the peak, and the ratio between its total flux and that
of the original MIPS PSF is 1.369.
For construction, this factor represents also the ratio between the equivalent areas of
the original PSF and the restored gaussian beam, in spite of having
the same FWHM. In other words, the Spitzer PSF affects a larger area
of the sky and suffers of a greater confusion.
The mentioned procedure allows us to obtain a map (the CLEAN map) where the
Airy rings are removed. 

An example of the effects of applying the CLEAN procedure is
illustrated in Figure \ref{map2}, where we report a zoom of 2.2$\times$1.4 
square arcminutes into the map presented
in Figure \ref{map1}. The upper panel shows the original flux map, in
the lower panel the corresponding cleaned map shows that the adopted
approach is efficient in deconvolving many close sources, providing
a reconstructed map where the confusion due to first Airy ring is reduced.
We will show that this procedure is not
sufficient to solve most of the confusion, and that a more accurate
approach to deal with these cases is needed.

\begin{figure*}
\psfig{file=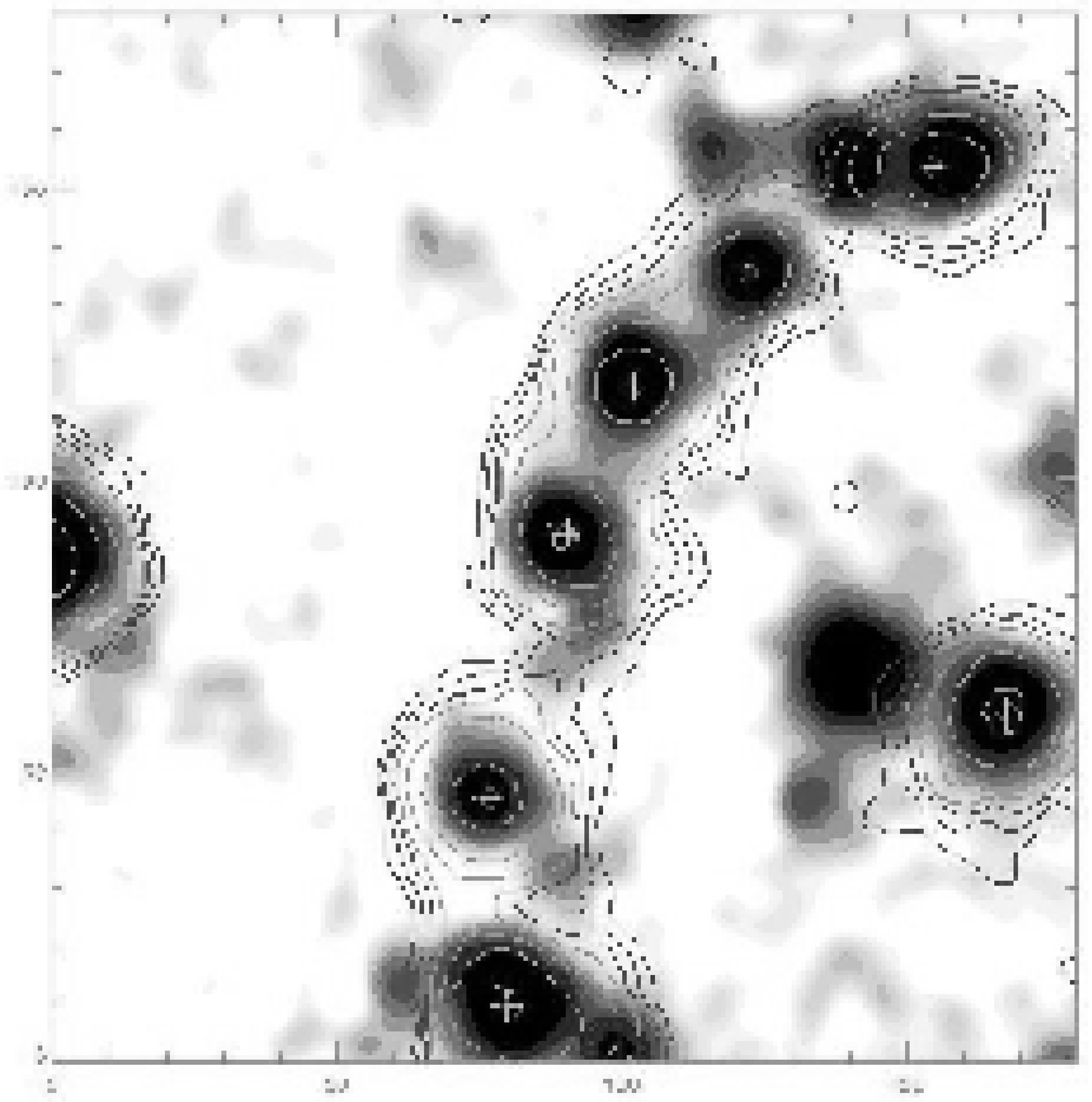,width=7.5cm}
\psfig{file=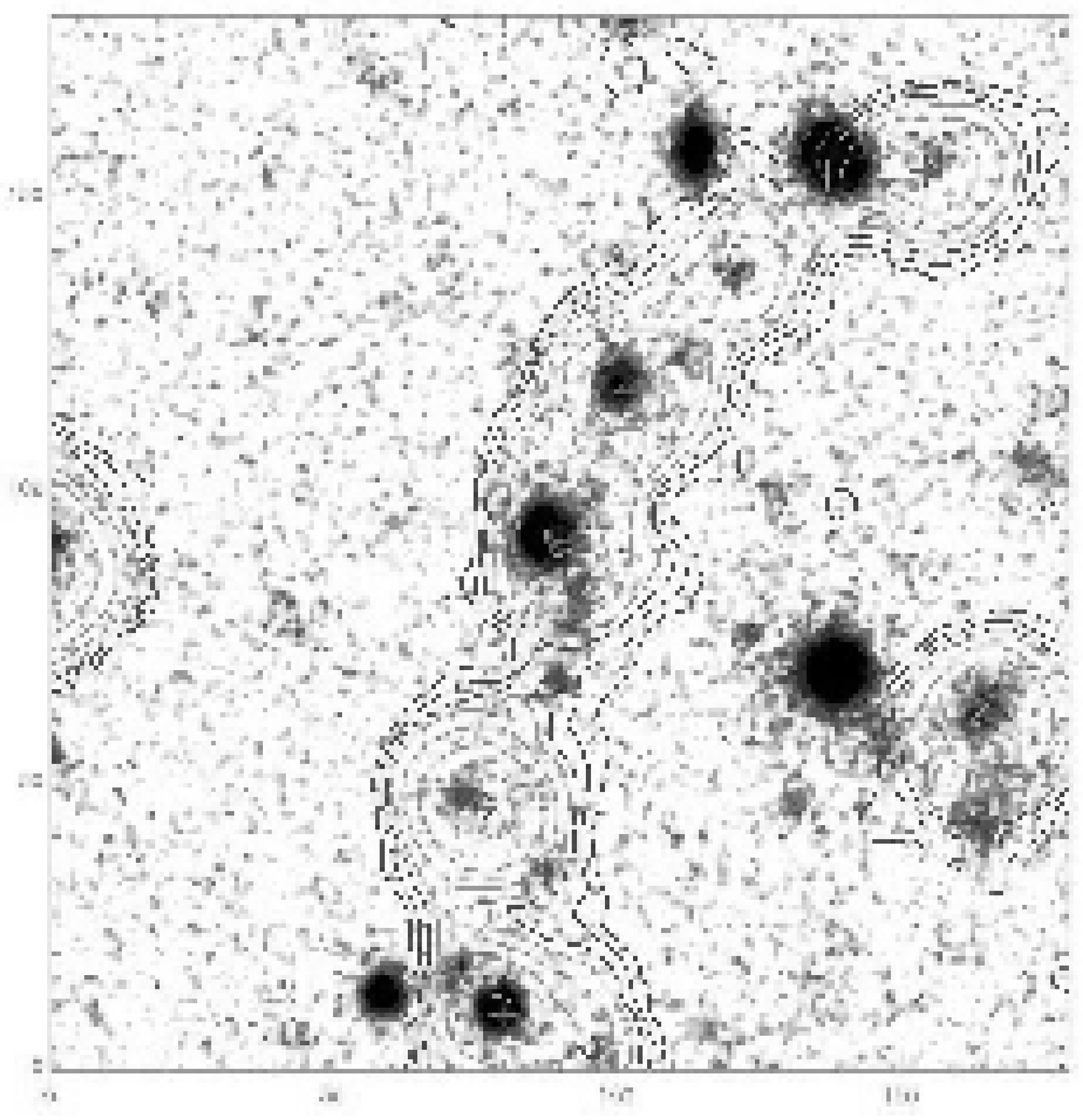,width=7.5cm}
\psfig{file=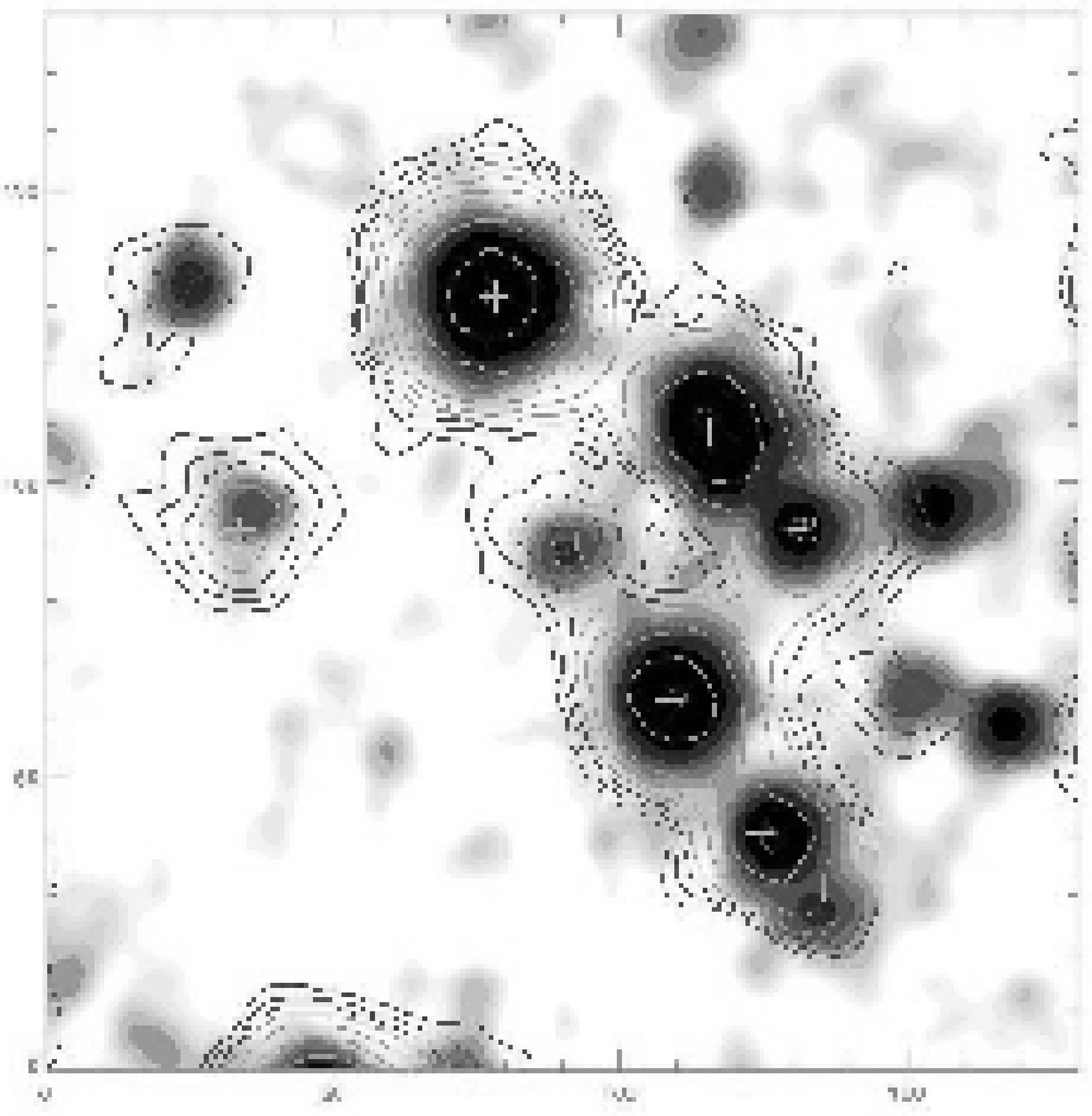,width=7.5cm}
\psfig{file=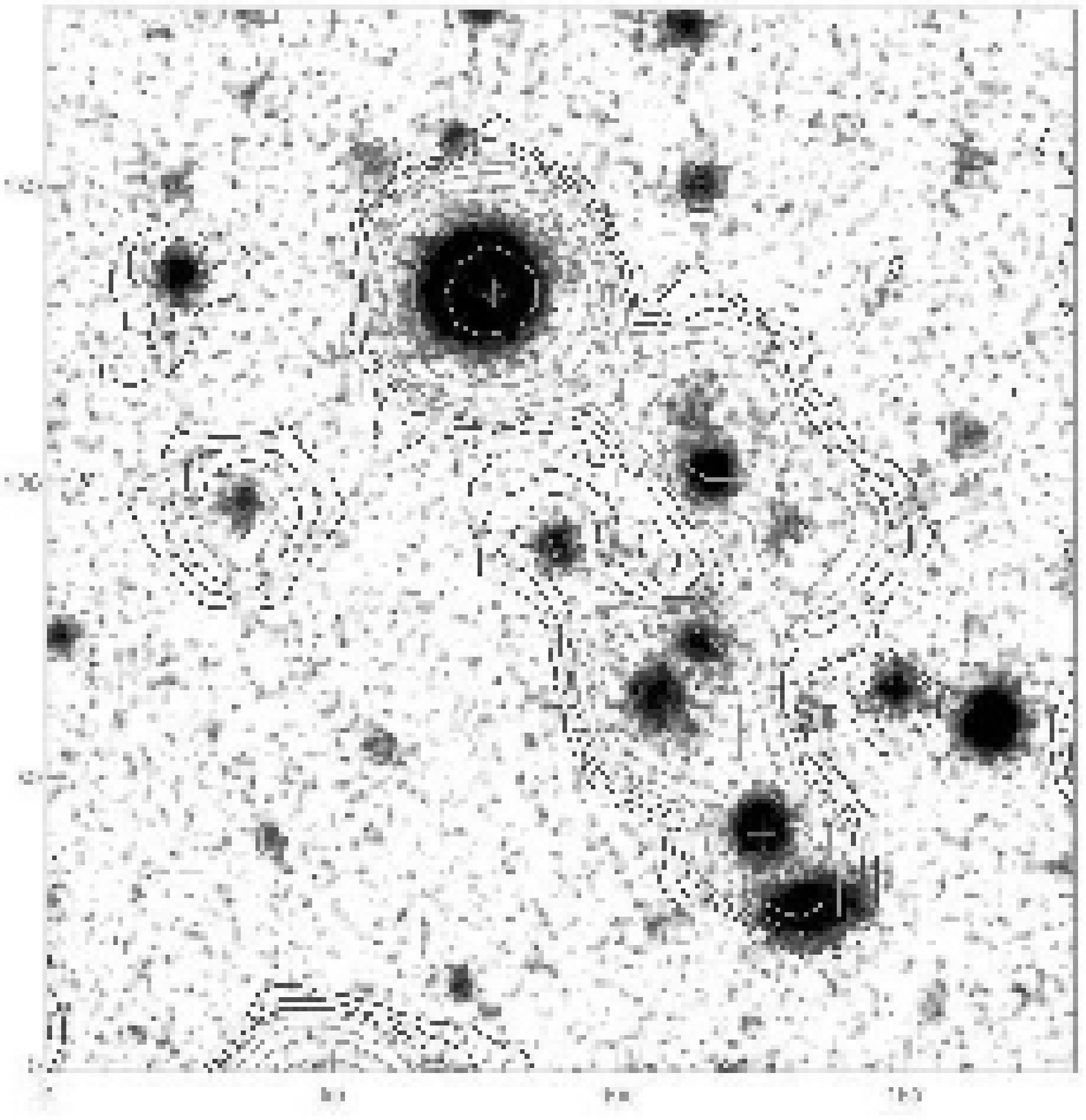,width=7.5cm}
\psfig{file=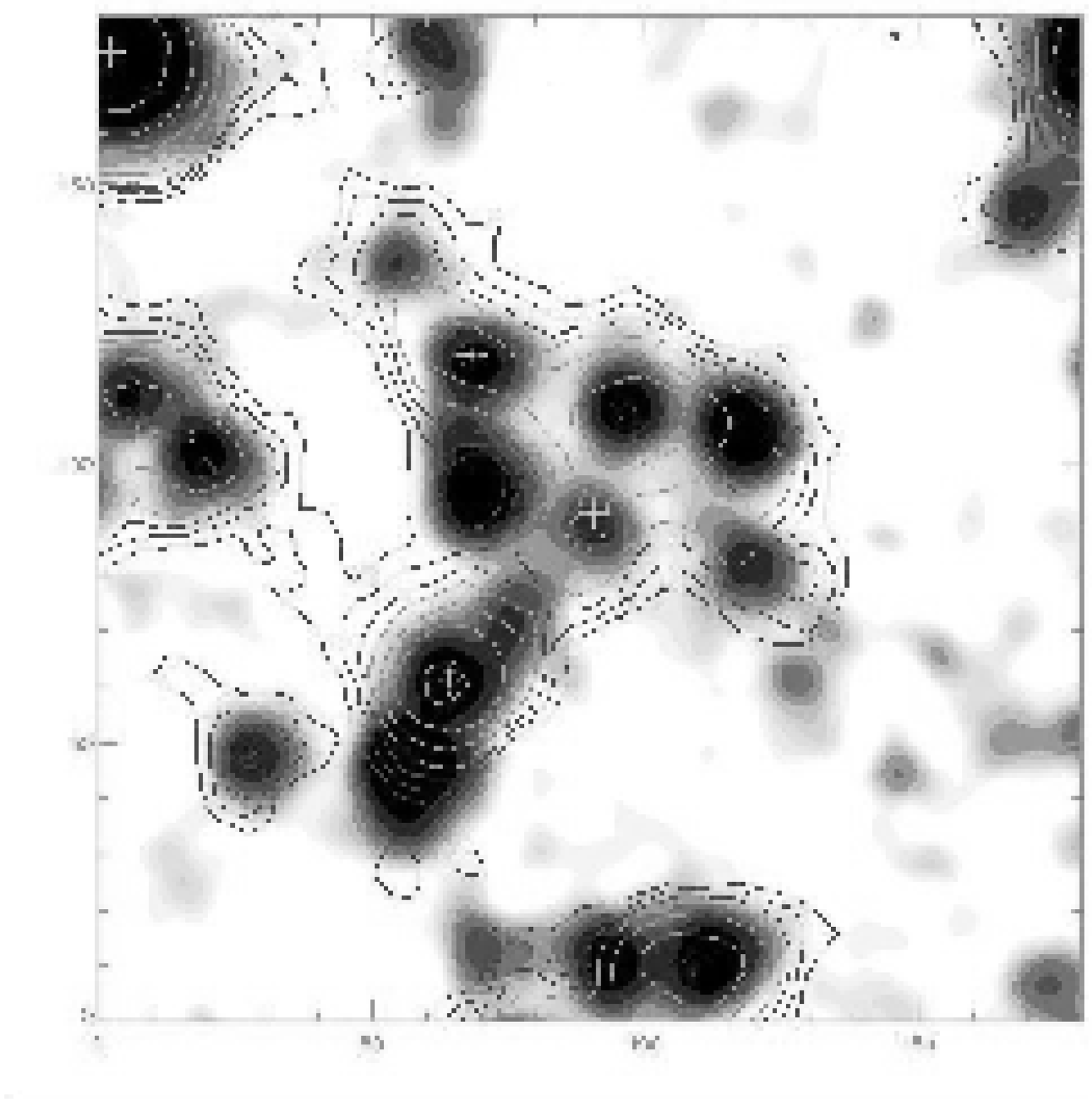,width=7.5cm}
\psfig{file=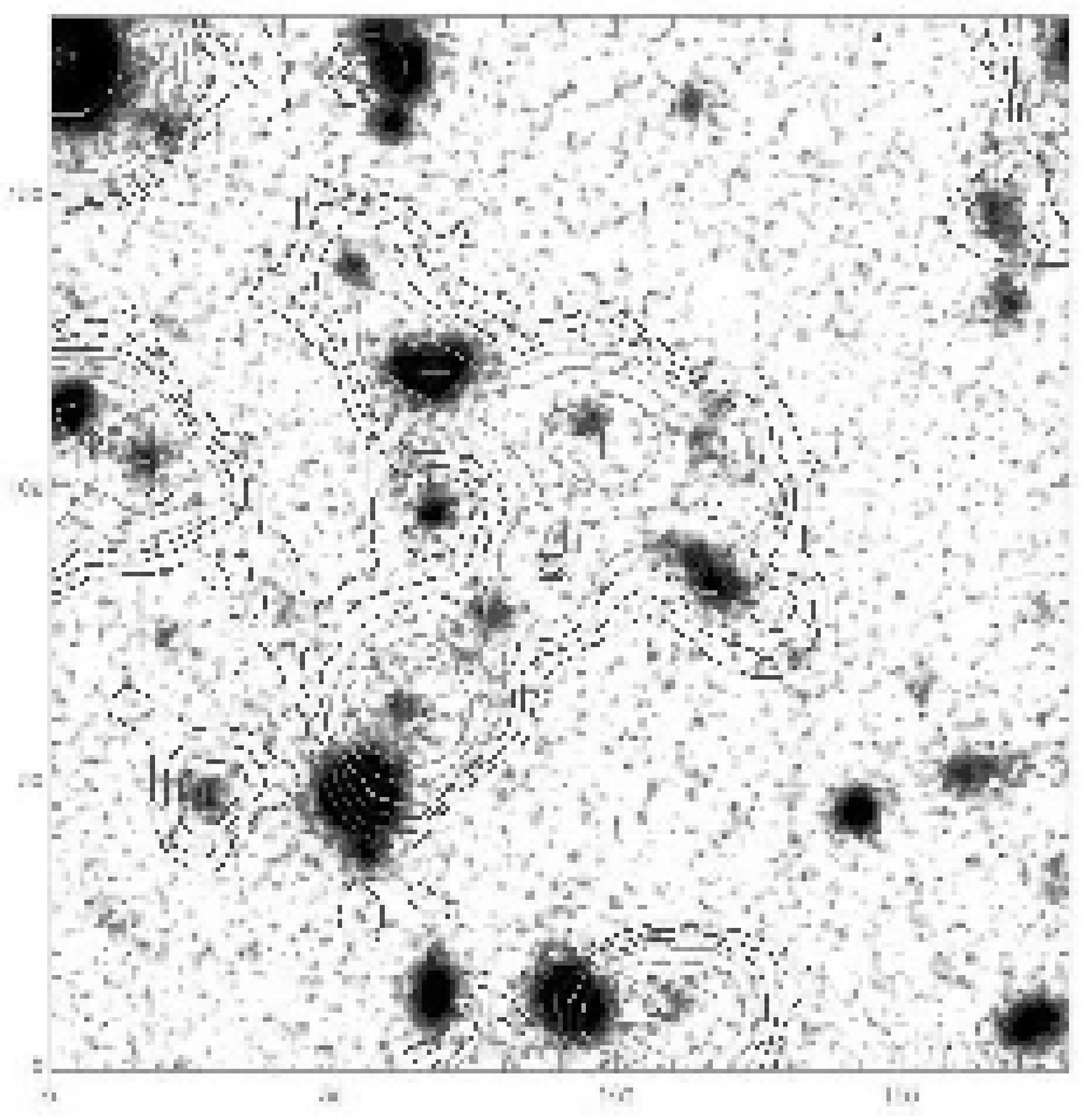,width=7.5cm}
\caption{
Example of three blend 24 $\mu$m sources. For each  case we
report both the near-IR (IRAC 3.6 $\mu$m, left panel) and the
optical (R band, right panel) images overlayed
with the MIPS 24 $\mu$m contours, starting from the 2-sigma level at increasing values.
North is up, East at left. Each image is 50$\times$50 arcsec.
The crosses mark the positions of the 24 $\mu$m sources in the catalogue  
with signal-to-noise ratio greater than 5.
}
\label{blend6}
\end{figure*}

\subsection{Source Extraction}
\label{se}
Before performing source extraction, we have applied a recursive median 
filtering to the final mosaic map. The size of the box used to
computed the median is 64$\times$64 pixels of 1.2 arcsec. This procedure
is used order to smooth residual spurious background fluctuations.
Source detection is then performed on the signal-to-noise map by 
selecting all pixels above a low flux threshold (0.5 $\mu$Jy pixel$^{-1}$) using the IDL 
Astronomy Users Library routine called $find$
(based on DAOPHOT's equivalent algorithm). We then extract from the
detection list only the objects with signal-to-noise ratio greater
than 5. The preliminary list includes 953 sources. 
The combination of long exposure times and high repetition factor of this Spitzer survey
provides a very deep infrared observation of the sky, that is basically confusion limited.
For this reason we checked each detected source by visual inspection.
We exploited the availability of higher frequency observations from IRAC (see section \ref{obs}) to look
for the counterparts of eventual blended MIPS sources. Indeed, the better resolution of the Spitzer
near-IR data allows one to disentangle the multiple components of the mid-IR sources.
We found that 21\% of 953 sources are ``blended''. These objects have generally pairs 
or small groups as NIR/optical counterparts.
A representative sample of this class is shown in Figure \ref{blend6}, where for each blend case we
report both NIR (IRAC 3.6, left panel) and 
optical (R band, right panel) images overlayed
with the MIPS 24 $\mu$m contours ($> 3-\sigma$).

\subsection{Photometry}
\label{photom}
Since most of the objects in our catalogue are faint point-like sources,
their 24 $\mu$m flux density is computed by applying a correction factor to
the measured peak flux, in order to convert the peak flux into a total flux density.
The correction factor is basically computed from the {\em Spitzer} MIPS PSF, 
provided within the Mopex package (as of June 2004), and is found to have a value of 32.5. However, other authors
report slightly different aperture corrections, which are taken into account
when comparing the source counts (see Section \ref{ecount}).
In this work the standard calibration from the SSC is applied. Future detailed measures of stellar fluxes,
to be compared with atmosphere model predictions, will provide a more robust absolute calibration
for Spitzer photometry.  The MIPS 24 $\mu$m absolute flux calibration is correct within few percent. 

The measure of the total flux from its peak value has been applied to every isolated source.
For the few extended sources (three) aperture photometry was computed.
A different approach is required in the case of confused and blended sources. 
We have examined each single MIPS blend source and determined the multi-counterparts looking
at the IRAC image. When the associations are not clear, we refer to the optical R band
frames that allow us to visually check and solve most of the ambiguous cases.
Once the number and the positions of the counterparts of a mid-IR blend source are fixed,
a PSF fitting algorithm (IMFIT, within the AIPS environment) was
applied to compute an accurate photometric deblending on the CLEAN
map. Blended sources are assumed to be point like sources.
This provides a measure of the 24 $\mu$m flux of each confused component.
At the end of this analysis the final MIPS sample in the GOODS EN1
field includes 983 sources detected above the $5 \sigma$ threshold.

\subsection{The 24 $\mu$m GOODS EN1 catalogue}
\label{catalogue}
The adopted deblending procedure turns out to be an efficient tool to
break through the confusion limit of the deep MIPS 24 $\mu$m
surveys, such as GOODS. The use of combined higher frequency maps
(near-IR and optical in our case) has allowed us to build an unbiased sample of
faint IR sources. This will be discussed in more detail in Section \ref{confusion}.

\begin{figure}
\psfig{file=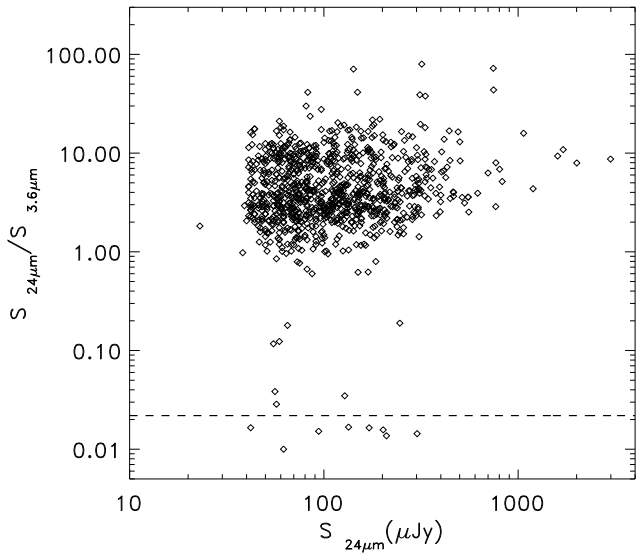,width=9cm}
\caption{$S(24\mu m)/S(3.6 \mu m)$ flux ratio as a
function of the 24 $\mu$m flux for the sources in the final
catalogue. The dashed horizontal line marks the
Rayleigh-Jeans ratio. We classified as stars all sources with
$S(24\mu m)/S(3.6 \mu m)< 0.1$.
}
\label{stars}
\end{figure}

As an illustration, in Table \ref{tab} we report a tabulation of the
first 10 rows of the final catalogue of 24 $\mu$m 
detections in the GOODS EN1 validation field. The complete
catalogue{\footnote {The complete catalogue in ASCII format is made
    publicly available through the world--wide-web:
    http://dipastro.pd.astro.it/giulia/EN1/ver or directly on
    request the authors.}}
 contains 983 sources 
detected above 5$\sigma$ and down to a minimum flux level of 23 $\mu$Jy. 
The five entries in the Table are: IAU source designation, right 
ascension and  declination (J2000), signal-to-noise ratio, total 24
$\mu$m flux density (in $\mu$Jy).
Fluxes marked on the right side with an asterisk indicate the 11
sources classified as stars. Figure \ref{stars} shows
the distribution of the $S(24 \mu m)/S(3.6 \mu m)$ flux ratio as a
function of the 24 $\mu$m flux for the sources in the final
catalogue. This near-to-mid IR colour is an efficient star/galaxy separator 
(a similar technique was already used by us for 15 $\mu$m ISOCAM sources,
Rodighiero et al. 2004). The dashed horizontal line marks the
Rayleigh-Jeans ratio. We classified as stars all sources with
$S(24\mu m)/S(3.6\mu m) < 0.1$.
The few points above the dashed line correspond to elliptical galaxies.

The flux uncertainties are basically given by the signal-to-noise ratio.

\section{Effects of confusion}
\label{confusion}
\begin{figure}
\centerline{
\psfig{file=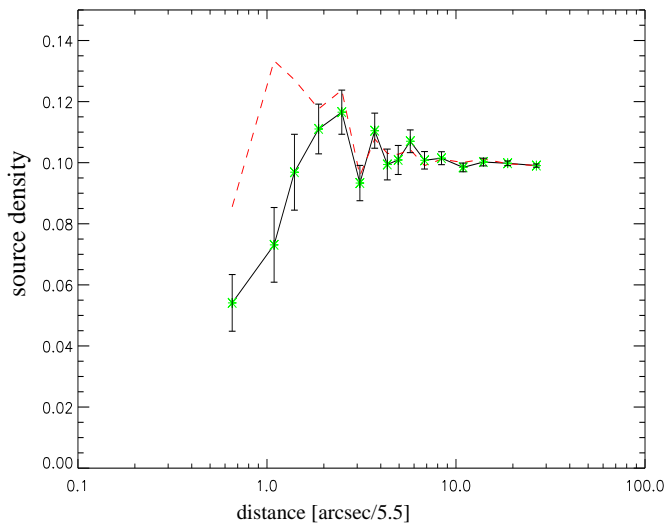,width=9cm}
}
\caption{
Number of detections falling
within a circle of radius $r$ (centered on the source position) 
normalized to a circular area with radius of 5.5'',  
as a function of distance $r$ for all the sources in the catalogue.
The solid line represents the preliminar 5-$\sigma$ catalogue 
before deblending the confused sources. 
The dashed line represents the final deblended catalogue.}
\label{pairs}
\end{figure}

To quantitatively estimate the global effect of confusion in the EN1 GOODS 24 $\mu$m map,
we have measured the density of galaxy pairs at increasing distance scale.
The result is shown in Figure \ref{pairs}.

\begin{figure}
\psfig{file=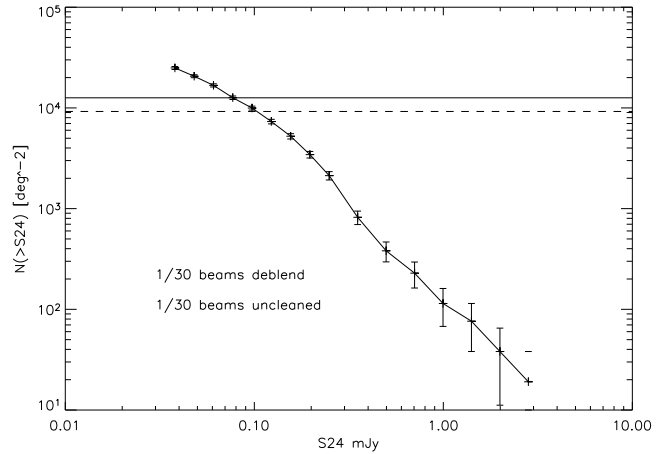,width=9cm}
\caption{
Final integral source counts (computed as
described in Section \ref{ecount} from the deblended catalogue). The horizontal lines mark the
level at which the classical confusion noise is reached (30 beams per
source, Condon et al. 1974). The solid line is computed 
by using the CLEAN gaussian beam, while the dashed line is derived
from the standard PSF beam.
}
\label{conf_beam}
\end{figure}

For each single source we have counted the number of detections falling
within a circle of radius $r$  centered on the source position. 
This is recursively done for different increasing
values of $r$, and for all the sources in the catalogue.
The median value of the source density at each distance $r$ is
normalized  to a circular area with radius of 5.5'',
and plotted against $r$.
The solid line in the figure is computed using the preliminar 5-$\sigma$ catalogue 
before deblending the confused sources (it includes 953 sources), while
the dashed line has been built from the final deblended catalogue (983 sources).
No corrections were applied for the variable noise, but only for the geometry as the larger circles are limited
by the boundary of the map.
This analysis indicates the completeness of our method: the only sources that we are not able to
deblend only those too close to each other (angular separation
lower than $\sim$ 6''). 
Obviously, identifications from higher resolution instruments
may help, but if the separation is too small, even IRAC observations might suffer 
of confusion problems.
A Montecarlo simulation test can recover these small deficits when dealing with source counts, 
however we have been very conservative when computing the source counts, cutting at a
flux level which is not affected by the mentioned problem.

We have directly measured the confusion level of the 24 $\mu$m observations
from the observed cumulative source counts. This is presented in
Figure \ref{conf_beam} 
where we show our final integral source counts (computed as
described in Section \ref{ecount} from the deblended catalogue). The horizontal lines mark the
level at which the classical confusion noise is reached (30 beams per
source, Condon et al. 1974). 
The solid line is computed by using the CLEAN gaussian beam, while the dashed line is derived
from the standard PSF beam (the latter being a factor of 1.369
greater than the former, see Section \ref{se}).
This result confirms that the approach of using higher frequency to
constrain the multi-counterparts of the blend/confused mid-IR sources is succesfull
in reducing the intrinsic confusion limit of deep mid-IR maps.
In particular, we have obtained 24 $\mu$m confusion limits of 77 $\mu$Jy
and 103 $\mu$Jy from the final and the pre-cleaned samples respectively. 
The confusion limit is then decreased of about 30\%.
These values are consistent with those extrapolated from model predictions
by Dole et al. (2004). They report three different estimates based on
different methods: 56 $\mu$Jy (from the source density criterion, Dole
et al. 2003), 71 and 141 $\mu$Jy (levels deduced from the source
density of one source per 20 and 40 independent beams).


\section{Extragalactic number counts}
\label{ecount}
To study the statistical properties of our sample, and to alternatively
check the effects of confusion on our results, we have computed the
extragalactic source number counts. Stars were 
removed from this compilation.
We have computed the 24 $\mu$m source counts down to a flux level of
$\sim$30 $\mu$Jy. The counts have been obtained by weighting each
single source for the effective area corresponding to its flux density.

\begin{figure*}
\centerline{
\psfig{file=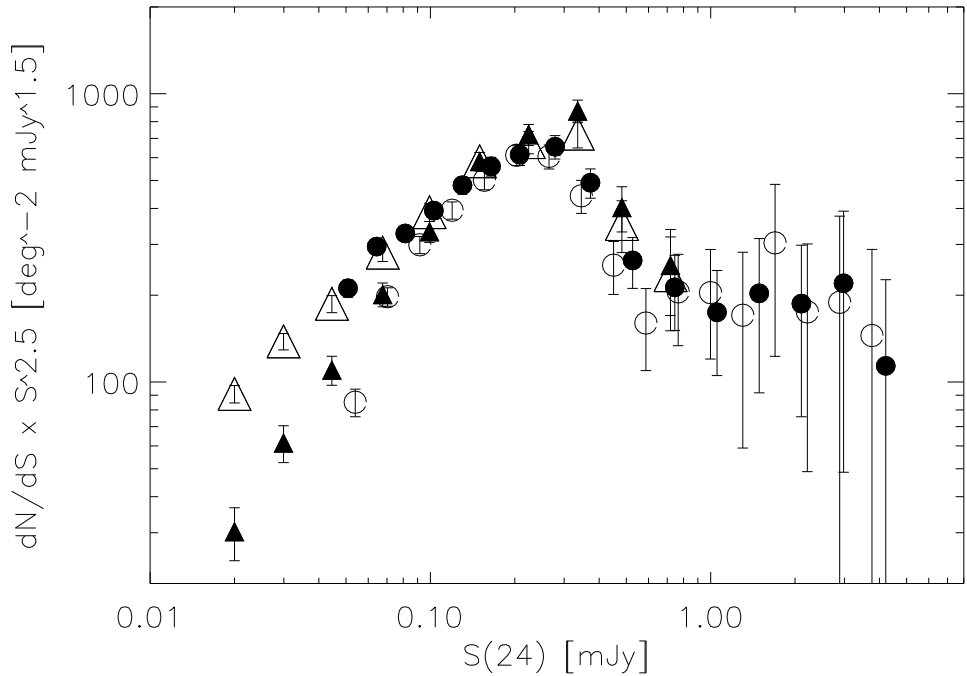,width=14cm}}
\caption{Differential 24 $\mu$m counts normalized to the Euclidean law
  ($N\propto S^{-2.5}$). 
We compare the
distribution obtained by using our final deblended catalogue 
(filled circles, 983 sources) and the preliminary 5-$\sigma$ catalogue prior to
deblending (open circles, 953 sources). We compare our results with the
source counts by Chary et al. (2004). The filled triangles refer
to their uncorrected counts, the open triangles correspond to
their final completeness corrected data.
}
\label{counts1}
\end{figure*}

In Figure \ref{counts1} we report the differential 24 $\mu$m counts
normalized to the Euclidean law ($dN/dS \propto S^{-2.5}$). We compare the
distribution obtained by using our final deblended catalogue 
(filled circles, 983 sources) and the preliminary 5-$\sigma$ catalogue obtained
before deblending (open circles, 953 sources).
The errors associated to the counts have been
computed as $\sqrt{\sum_i 1/A^2_{eff}(S_i)}$, where the sum is for all
the sources with flux density $S_i$ and $A_{eff}(S_i)$ is the
effective area corresponding to that flux. The quoted errors have to
be considered as lower limits of the total errors.

The two distributions reported in Figure \ref{counts1}
clearly show that, when accounting for confusion,
the density of faint sources is generally increased, and the source
counts slope is less steep (by a factor of $\sim$2) below the peak observed at $\sim 0.2$ mJy.

In section \ref{confusion} we argued that an accurate deblending
analysis might provide un-biased complete samples. This is
quantitatively confirmed if we compare our results with those derived
by Chary et al. (2004), starting from the same data-set in the EN1
GOODS 24 $\mu$m field, while they have not treated each single source by
looking at the counterparts in the IRAC map. They have followed a
statistical approach to study the properties of the 24 $\mu$m
population. Their completeness corrections were measured
using a Monte Carlo approach: by adding artificial sources on
the original mosaic, they were able to recover the fraction of
undetected/lost sources as a function of the flux level combined with
their extraction procedure.
In Figure \ref{counts1} we report also the
source counts by Chary et al. (2004) for comparison with our estimate. 
The filled triangles refer
to their uncorrected counts, while the open ones correspond to
their final completeness corrected data.

We have mentioned (Section \ref{photom}) that to compare our data with
those from other authors we need to apply a correction factor to
report the different results to the same flux scale. 
For consistency, we have firstly scaled upward by a factor of 1.057
the fluxes of our sources. This factor is needed to recover the light
lost in the outer PSF lobes, and it has been computed as the
ratio between the integral of the PSF and the flux of the same
PSF measured within an aperture of 37.4''.
We have not applied this correction to the fluxes presented in the
final catalog, as it has been derived from a theoretical PSF and
should instead be better constrained from an absolute calibration based
on atmospheric stellar emission models compared to the observations.
Chary et al. (2004) have computed the flux of a source in a circular aperture
of radius 6'' and corrected upward by a factor of 1.8 to account for
the wings of the PSF. Using the PSF derived in this paper (see Section \ref{photom})
we found that the factor to convert 6'' aperture fluxes to total fluxes
is about 1.45. This implies that we need to correct the fluxes in the source counts 
reported by Chary et al. (2004) downwards by a factor
1.8/1.45=1.24 to compare them with our estimates on the same scale.
A reason for this discrepancy may rely on the PSF provided by Mopex
(at least for the version of June 2004 used in this work),
which is not extended enough to recover the total flux lost in its wide
wings. 
This is confirmed by testing the correction factor at the same
aperture considered by Chary et al., 6'', using a different PSF,
computed with Spitzer/TINYTIM by Fadda et al (2005, in prep.)
and that is quite similar to that measured from a real star in
the First Look Survey (D. Fadda 2005, private communication).
The aperture correction in this case is 1.76, very close to the 1.8
factor reported by Chary et al. (2004).


Once the flux scale correction is applied, the comparison between Chary et al. counts and
ours shows a remarkable consistency (see Figure \ref{counts1}). 
This result is a direct demonstration that we do not need 
to apply any statistical correction to our sample, which turns out to
be almost un-biased down to relatively faint fluxes (40 $\mu$Jy).

A major advantage of having carefully accounted for the contribution of
confused sources, is that we have produced a catalogue which is not only
suited for statistical purposes, but also to study the spectro-photometric 
properties of the single faint mid-IR sources.

\begin{figure*}
\centerline{
\psfig{file=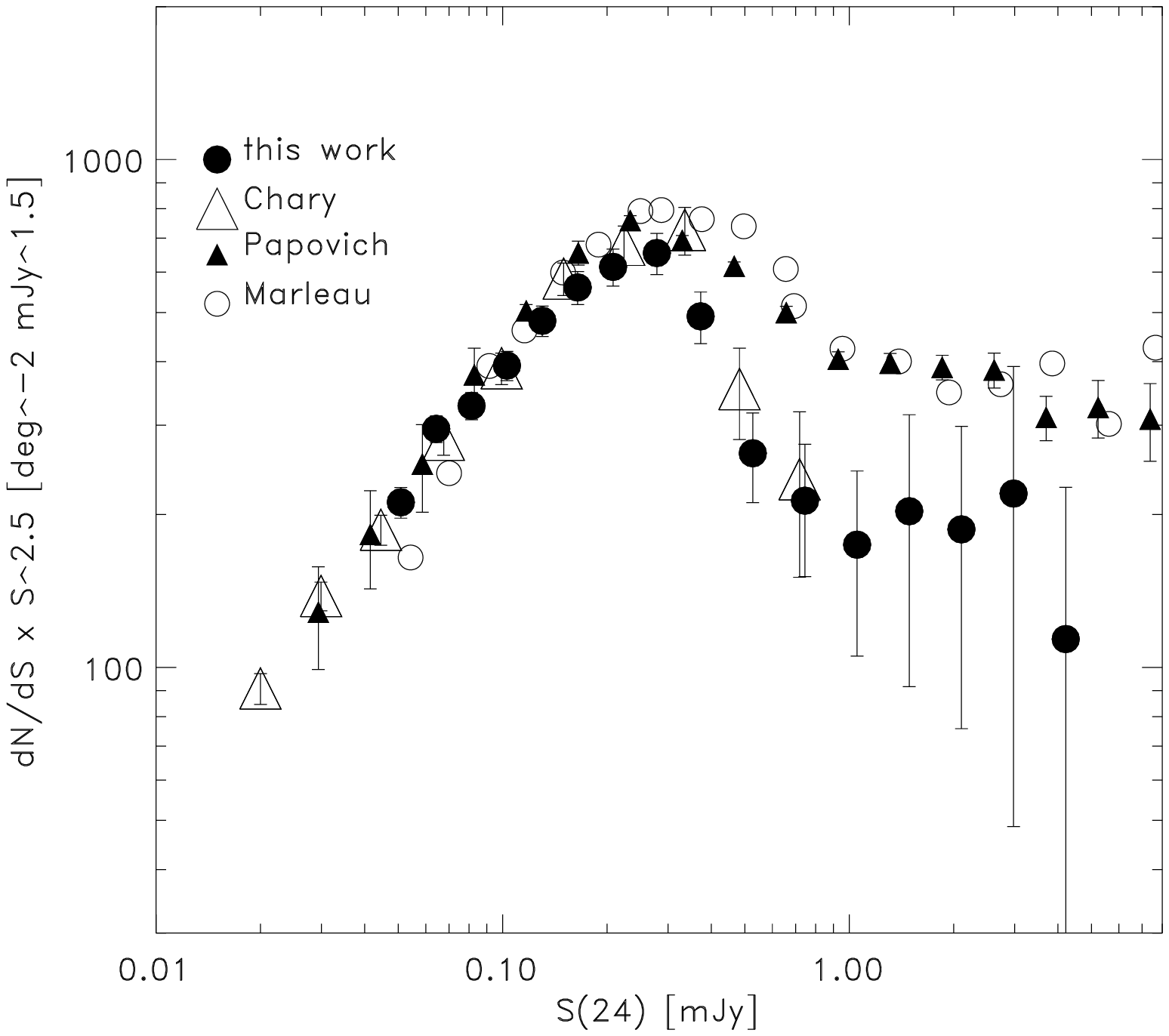,width=14cm}}
\caption{Differential 24 $\mu$m counts normalized to the Euclidean
law. We compare our source counts (filled circles) with those of
Chary et al. (2004, open  triangles), Papovich et al. (2004, filled
triangles) and Marleau et al. (2004, open circles).
}
\label{counts2}
\end{figure*}

In Figure \ref{counts2} in addition to comparing our source counts (filled circles) to 
those of Chary et al. (2004, open triangles), we show also those from Papovich et
al. (2004, filled triangles).
As for the case of Chary et al counts, we have applied a correction factor
to the data of Papovich et al. (2004). Their source photometry
corresponds to the flux of a PSF (derived from the bright stars in the field)
within an aperture diameter of $37^{''}.4$, then they apply a multiplicative
correction of 1.14 to account for light lost outside this
aperture. From our ``edge cut'' PSF we find that a 37$^{''}.4$ aperture flux is a
good measure of the total flux. This implies that we need to correct downward by a factor
1.14/1.0=1.14 the fluxes in the source counts of Papovich et al. (2004) to bring
them on the same scale as ours.

We also report the 24 $\mu$m counts published by Marleau et al. (2004,
open circles). 
They use a correction of 1.1 (D. Fadda, priv. comm.) to
compute the total flux from an aperture photometry of radius 15.24''. 
At the same aperture we compute a correction factor of 1.09.
The scaling factor to be applied to Marleau et
al. counts, in order to have them referred to our flux scale, is
then very small, 1.1/1.09=1.01. 

Although Figure \ref{counts2} shows a general agreement
between the four independent 24 $\mu$m counts analysis 
over about a decade in flux (at low fluxes: 0.03 mJy -- 0.3 mJy), at fluxes brighter 
than 0.3 mJy the GOODS N1
statistics is quite poor as the field was explicitly choosen to avoid
bright mid-IR sources
(based on the positions of ISOCAM sources).
The study of Papovich et al. (2004) is based on a much wider area
($\sim$10 deg$^2$), splitted in different sky regions. Poor statistics and cosmic variance
on small scales (as the size of the EN1 field) can be the reason for 
the counts discrepancy observed above 0.3 mJy.

By integrating our source counts down to S(24 $\mu$m)= 38 $\mu$Jy
we have obtained an estimate of the 24 $\mu$m cosmic background intensity
down to that flux level: 1.25 $nW m^{-2} sr^{-1}$. We have then extrapolated the 
differential source counts using a power law 
fitting the fainter flux bins (slope = -0.58) to obtain an estimate of the
expected value of the total background at 24 $\mu$m: 2.016
$nW m^{-2} sr^{-1}$. The fraction of the background that we 
resolve down to 38 $\mu$Jy is about 62\%.
Papovich et al. (2004) report a value for the total background intensity at
24 $\mu$m of 2.7 $nW m^{-2} sr{-1}$, a factor of $\sim 30\%$ higher than our estimate.

\section{Summary}
\label{summary}
We have analysed a deep MIPS 24 $\mu$m survey in the EN1 region,
conceived as a validation observation for the official GOODS strategy
(Dickinson et al. 2004). The archival Spitzer data of the GOODS EN1
test field allowed us to check the effects of source confusion on the mid-IR statistics.
We have processed the data to obtain a final mosaic that is basically
confusion limited. To deal with confusion, we have applied the CLEAN
algorithm to the 24 $\mu$m map, removing the diffraction rings around bright
sources. Few blend sources are then partially recovered and appear
as resolved objects in the CLEANed map.
A different approach is needed for confused sources with closer single counterparts. We
used the information from higher frequency observations in the same
regions in order to detect the counterparts of the apparently extended
24 $\mu$m blend sources. An optimal solution is provided by the
near-IR camera on board Spitzer, IRAC. In particular, we have referred
to the 3.6 $\mu$m and 4.5 $\mu$m maps to associate the single counterparts to
the sub-components of 
the 24 $\mu$m unresolved emission. An optical R-band image (Fadda et
al. 2004) has also been considered for disentangling a few dubious cases. 
We have then obtained a reliable 24 $\mu$m catalogue including 983 sources at the
5-$\sigma$ level, the fainter source reaching a flux of 23 $\mu$Jy.
We have shown that our sample is almost unbiased down to $\sim$40 $\mu$Jy,
and that by applying a deblending procedure we do not need to apply
any statistical completeness correction (at least to the flux level considered).
This is probed by direct comparison of our results with those of Chary
et al. (2004), who analysed the same data-set through extensive Monte
Carlo simulations.

We have estimated the confusion limit of the MIPS 24 $\mu$m survey
using the classical level definition of 30 independent beams per source,
obtaining a value of 77 $\mu$Jy. We also found that the deblending
procedure has reduced of about 30\% the confusion noise.

We have compared the 24 $\mu$m extragalactic source counts derived
from our sample with those published by Chary et al. (2004) and
Papovich et al. (2004). We found a good agreement between the
different data-sets, when reporting the counts of the 
other authors to the same flux scale. At the bright end, the GOODS EN1
test field is biased by low statistics and probably cosmic variance,
resulting in an underdensity of bright sources at $> 0.3$ mJy
with respect to larger area surveys (Papovich et al. 2004).

By extrapolating the source counts to fainter fluxes, we have
estimated the fraction of the cosmic background already
resolved by our sample down to a flux level of 38 $\mu$Jy: $\sim$62\%.

The main goal of this work was to provide a reliable catalogue of 
mid-IR sources. This is essential in order to study the physical
nature of the fainter population responsible for most of the cosmic IR
background. 


\begin{table}
 \caption[]{First 10 entries of the 24 $\mu$m source catalogue in the
 EN1 test field}
\label{tab}
\begin{tabular}{c c c c c}
   \hline
     ID &  RA          & DEC           & S/N   & S(24$\mu$m)\\
      ~ &~&~&~& [$\mu$Jy]\\
   \hline
EN1-160855+550112 & 16:08:55.87  & 55:01:12.05   & 194& 2990\\
EN1-160942+550009 & 16:09:42.66  & 55:00:09.48   & 191& 1597\\
EN1-160906+545557 & 16:09:06.53  & 54:55:57.38   &  94&  747\\
EN1-160904+545821 & 16:09:04.47  & 54:58:21.75   &  94& 1067\\
EN1-160848+545150 & 16:08:48.87  & 54:51:50.46   &  92& 1194\\
EN1-160936+550242 & 16:09:36.91  & 55:02:42.39   &  69&  557\\
EN1-160921+545109 & 16:09:21.50  & 54:51:09.93   &  62&  700\\
EN1-160839+545523 & 16:08:39.85  & 54:55:23.39   &  62&  553\\
EN1-161003+545344 & 16:10:03.27  & 54:53:44.77   &  57&  803\\
EN1-160839+545523 & 16:08:39.85  & 54:55:23.39   &  54&  553\\
   \hline
   \hline
\end{tabular}
\end{table}
\end{document}